# Optimal design of a bilayer for the highest thermal resistance: A lesson learned from the shells of snails from hydrothermal extreme environment

Anran Wei, Haimin Yao*

Department of Mechanical Engineering, The Hong Kong Polytechnic University, Hung Hom, Kowloon, Hong Kong, China

* Corresponding author, E-mail: mmhyao@polyu.edu.hk

**Abstract**

Inspired by the unique design of the shells of snails inhabiting the deep-sea hydrothermal environment, here we theoretically study the temperature response of a bilayer to an external thermal impulse. A semi-analytical solution to the temperature field in the bilayer is obtained, allowing us to assess the peak temperature that occurs on the inner wall as a quantitative indicator of the thermal resistance of the bilayer. The structural determining factors of the thermal resistance of a bilayer are then investigated by examining the effects of the stacking sequence and volume fractions of the constitutive layers on the peak temperature on the inner wall. Our results indicate that the stacking sequence of the two layers in a bilayer, as well as their volume fractions, play important roles in determining the thermal resistance. For two layers with given materials, there exists an optimal stacking sequence and thickness ratio giving rise to the best thermal resistance. The results of our work not only account for the unique laminated design of the snail shells from hydrothermal environments but also provide practical guidelines for the design of multilayer thermal barriers in engineering.

# 1. Introduction

Materials and structures applied in extreme environments tend to be subjected to various threats including high-temperature impulse, mechanical impact, and corrosion, and so on. Protection of these structures requires shields with not only good thermal resistance but also capabilities of bearing loading [1], anti-corrosion [2], and self-healing [3-5]. How to design and optimize a multifunctional thermal barrier remains a challenge in engineering. Studies on the protective exoskeletons applied in the natural extreme environments may bring us instructive inspirations.

One of the most typical extreme environments in nature might be the deep-sea hydrothermal vents (Fig. 1a). It has been unveiled that some gastropods inhabiting the deep-sea hydrothermal environment have evolved exoskeletal shells with a multilayered structure. For instance, Yao *et al*. found that the shell of *Crysomallon squamiferum*, a snail discovered in the Kairei hydrothermal vent field, shows a tri-layered structure with a highly calcified inner layer, an organic periostracum middle layer and a very thin iron sulfide-based outer layer [6], as shown in Fig. 1b. Similarly, a bi-layered structure is observed in the shell of *Alviniconcha hessleri* [7], another species of gastropod living near the deep-sea hydrothermal vent, as shown in Fig. 1c. A common structural feature in both snail shells is that a completely organic layer is deployed outside the calcified ceramic layer. Such layout of the organic and inorganic materials is quite different from that observed in the shells of land snails and regular marine gastropods (*e.g.*, abalone), in which the organic and inorganic ingredients are hybridized to form micro/nanocomposites. It is believed that such a unique layout of the organic and inorganic materials might be the consequence of adaption to the hydrothermal environment. A question thus arises. Given materials, how can we maximize the thermal resistance by tuning the deployment of the materials?

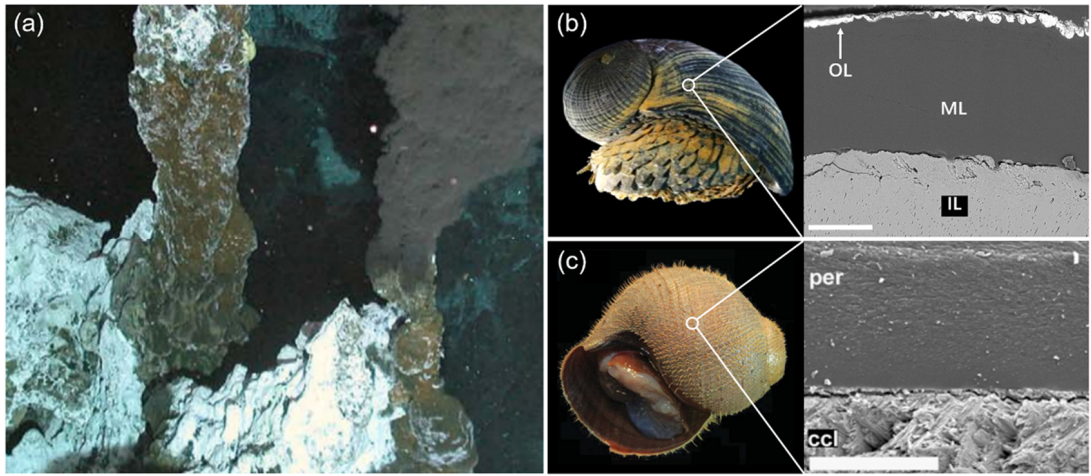

**Fig. 1.** (**a**) Photograph of an active deep-sea hydrothermal vent [8]. (**b**) Cross-section of the shell of the snail *Crysomallon squamiferum* living in the deep-sea hydrothermal environment (Adapted with permission from [6].) The IL, ML, and OL denote the highly calcified inner layer (IL), organic periostraca middle layer (ML), and iron sulfide-based outer layer (OL), respectively. The scale bar, 100 μm. (**c**) Cross-section of the shell of *Alviniconcha hessleri* [9, 10]. The PER and CCL denote the organic periostraca outer layer and complex crossed mineral lamellar inner layers, respectively. The scale bar, 100 μm.

Structural optimization has long also been applied to tune the thermal behavior of layered thermal barrier. For example, Cao *et al*. investigated the optimal layer thickness for the tri-layered thin-walled structure by topology optimization method [11]. However, the effect of the spatial distribution of individual compositions on the overall thermal resistance is still unclear. To gain an in-depth understanding of the unusual shell design of the hydrothermal snails and enhance the performance of the engineering thermal barriers, here we investigate the temperature response of a bilayer structure to an external high-temperature impulse.

## 2. Theoretical formulation

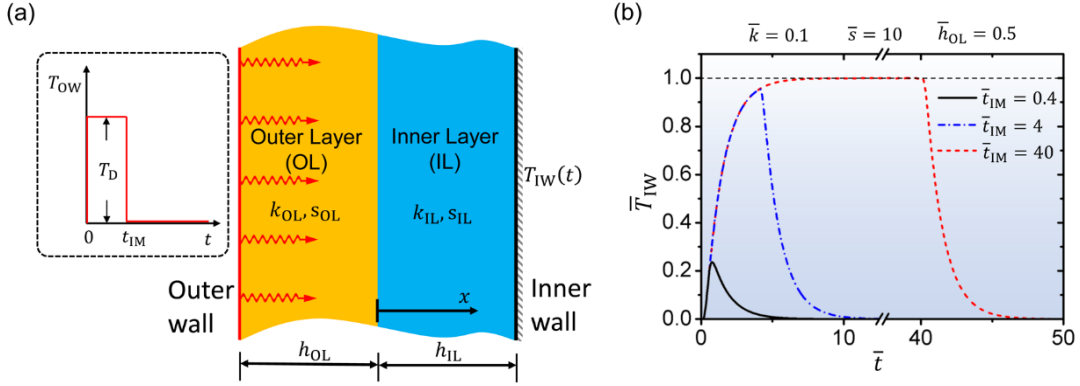

**Fig. 2** (**a**) Schematic depiction of 1-D thermal conduction model of a bilayer subjected to thermal impulse. (**b**) Calculated evolution of the temperature on the inner wall, $\bar{T}_{\mathrm{IW}}$, with the time $\bar{t}$ in response to the external thermal impulse with periods $\bar{t}_{\mathrm{IM}} = 0.4, 4, 40$, respectively. As an example, here the ratios of the thermal conductivity ratio and volumetric heat capacity between two layers are assumed as $\bar{k} = 0.1$ and $\bar{s} = 10$, respectively, and the thickness of the two layers are assumed the same (*i.e.* $\bar{h}_{\mathrm{OL}} = 0.5$).

Consider a bilayer structure composed of the outer layer (OL) and the inner layer (IL) with dissimilar thicknesses ($h$), thermal conductivities ($k$), and volumetric heat capacities ($s$), as shown in Fig. 2a. Initially, the whole system is at the temperature $T_0$. At $t = 0$, an instant temperature increment of $T_D$ is applied on the outer wall of the bilayer and lasts for a period of $t_{\mathrm{IM}}$, simulating the impact of an instantaneous thermal impulse. The inner wall of the bilayer is assumed thermally insulative and the thermal resistance of the interface between the OL and IL is neglected. The time-dependent temperature field in the bilayer, which is denoted by $T(x,t)$, should satisfy the governing equations of thermal conductivity as follows

$$\begin{cases} \dfrac{\partial T}{\partial t} = \dfrac{k_{\mathrm{OL}}}{s_{\mathrm{OL}}}\dfrac{\partial^2 T}{\partial x^2}, & (-h_{\mathrm{OL}} < x < 0) \\ \dfrac{\partial T}{\partial t} = \dfrac{k_{\mathrm{IL}}}{s_{\mathrm{IL}}}\dfrac{\partial^2 T}{\partial x^2}, & (0 < x < h_{\mathrm{IL}}) \end{cases} \qquad (1)$$

where $k_{\mathrm{OL(IL)}}$ and $s_{\mathrm{OL(IL)}}$ stand for the materials' *thermal conductivities* and *volumetric heat capacities* of the OL and IL, respectively.

Introducing dimensionless parameters

$$\bar{T} \equiv \frac{T-T_0}{T_\mathrm{D}}, \quad \bar{x} \equiv \frac{x}{h_{\mathrm{OL}}+h_{\mathrm{IL}}}, \quad \bar{t} \equiv \frac{t}{(h_{\mathrm{OL}}+h_{\mathrm{IL}})^2}\sqrt{\frac{k_{\mathrm{OL}}k_{\mathrm{IL}}}{s_{\mathrm{OL}}s_{\mathrm{IL}}}},$$

$$\bar{k} \equiv \frac{k_{\mathrm{OL}}}{k_{\mathrm{IL}}}, \bar{s} \equiv \frac{s_{\mathrm{OL}}}{s_{\mathrm{IL}}}, \bar{h}_{\mathrm{OL(IL)}} \equiv \frac{h_{\mathrm{OL(IL)}}}{h_{\mathrm{OL}}+h_{\mathrm{IL}}} \qquad (2)$$

Eq. (1) can be rewritten in a normalized form as

$$\begin{cases} \frac{\partial \bar{T}}{\partial \bar{t}} = \sqrt{\frac{\bar{k}}{\bar{s}}}\frac{\partial^2 \bar{T}}{\partial \bar{x}^2}, & (-\bar{h}_{\mathrm{OL}} < \bar{x} < 0) \\ \frac{\partial \bar{T}}{\partial \bar{t}} = \sqrt{\frac{\bar{s}}{\bar{k}}}\frac{\partial^2 \bar{T}}{\partial \bar{x}^2}, & (0 < \bar{x} < \bar{h}_{\mathrm{IL}}) \end{cases} \qquad (3)$$

The initial condition and boundary conditions can be given in a normalized form as

$$\bar{T}(\bar{x},0) = 0, \; \bar{T}(-\bar{h}_{\mathrm{OL}},\bar{t}) = 1 - \mathrm{H}(\bar{t}-\bar{t}_{\mathrm{IM}}), \; \frac{\partial \bar{T}(\bar{h}_{\mathrm{IL}},\bar{t})}{\partial \bar{x}} = 0 \qquad (4)$$

where $\bar{t}_{\mathrm{IM}} \equiv \frac{t_{\mathrm{IM}}}{(h_{\mathrm{OL}}+h_{\mathrm{IL}})^2}\sqrt{\frac{k_{\mathrm{OL}}k_{\mathrm{IL}}}{s_{\mathrm{OL}}s_{\mathrm{IL}}}}$ is the normalized duration of the thermal impulse, and $\mathrm{H}(\bar{t}-\bar{t}_{\mathrm{IM}})$ stands for a unit step function taking 0 when $\bar{t} < \bar{t}_{\mathrm{IM}}$ and 1 when $\bar{t} \geq \bar{t}_{\mathrm{IM}}$.

The continuity of temperature and conservation of heat flux across the interface between the OL and IL ($x = 0$) requires that

$$\bar{T}(0^-,\bar{t}) = \bar{T}(0^+,\bar{t}), \; \bar{k}\frac{\partial \bar{T}(0^-,\bar{t})}{\partial \bar{x}} = \frac{\partial \bar{T}(0^+,\bar{t})}{\partial \bar{x}} \qquad (5)$$

To solve the above partial differential equations (PDEs) about $\bar{T}(\bar{x},\bar{t})$, Laplace transformation is applied to Eqs. (3-5). Then the governing equations are converted into ordinary differential equations (ODEs) as follows:

$$\begin{cases} \frac{\partial^2 U}{\partial \bar{x}^2} - p\sqrt{\frac{\bar{s}}{\bar{k}}}U = 0, & (-\bar{h}_{\mathrm{OL}} < \bar{x} < 0) \\ \frac{\partial^2 U}{\partial \bar{x}^2} - p\sqrt{\frac{\bar{k}}{\bar{s}}}U = 0, & (0 < \bar{x} < \bar{h}_{\mathrm{IL}}) \end{cases} \qquad (6)$$

where function $U(\bar{x},p)$ is the Laplace transform of $\bar{T}(\bar{x},\bar{t})$, namely $U(\bar{x},p) = \mathcal{L}[\bar{T}(\bar{x},\bar{t})] = \int_0^\infty \bar{T}(\bar{x},\bar{t})e^{-p\bar{t}}d\bar{t}$. The corresponding boundary conditions and continuity requirements are also mapped into the complex domain as

$$U(-\bar{h}_{OL}, p) = \frac{1}{p}\left(1 - e^{-\bar{t}_{IM}p}\right), \quad \frac{\partial U(\bar{h}_{IL},p)}{\partial \bar{x}} = 0 \tag{7}$$

$$U(0^-, p) = U(0^+, p), \quad \bar{k}\frac{\partial U(0^-,\bar{t})}{\partial \bar{x}} = \frac{\partial U(0^+,\bar{t})}{\partial \bar{x}} \tag{8}$$

Solving Eq. (6) in combination with the conditions given by Eqs. (7-8) for function $U(\bar{x},p)$ and then taking the inverse Laplace transform give rise to the solution to $\bar{T}(\bar{x},\bar{t})$ as

$$\bar{T} = \begin{cases} \mathcal{L}^{-1}\left[\dfrac{2\left(1-e^{-\bar{t}_{IM}p}\right)\left(m_1 m_2 + m_3 m_4\sqrt{\bar{k}\bar{s}} - m_4 F\sqrt{\bar{k}\bar{s}}\right)}{p\left(m_1^2 m_2 + m_1 m_3 m_4\sqrt{\bar{k}\bar{s}}\right)}\right], & (-\bar{h}_{OL} < \bar{x} < 0) \\ \mathcal{L}^{-1}\left[\dfrac{2\sqrt{\bar{k}\bar{s}}\left(1-e^{-\bar{t}_{IM}p}\right)G}{p\left(m_1 m_2\sqrt{\bar{k}\bar{s}} + m_3 m_4\right)}\right], & (0 < \bar{x} < \bar{h}_{IL}) \end{cases} \tag{9}$$

where

$$\begin{cases} m_1 = e^{-\bar{h}_{OL}\sqrt{p\sqrt{\frac{\bar{s}}{\bar{k}}}}} + e^{\bar{h}_{OL}\sqrt{p\sqrt{\frac{\bar{s}}{\bar{k}}}}} \\ m_2 = e^{-\bar{h}_{IL}\sqrt{p\sqrt{\frac{\bar{k}}{\bar{s}}}}} + e^{\bar{h}_{IL}\sqrt{p\sqrt{\frac{\bar{k}}{\bar{s}}}}} \\ m_3 = e^{\bar{h}_{OL}\sqrt{p\sqrt{\frac{\bar{s}}{\bar{k}}}}} - e^{-\bar{h}_{OL}\sqrt{p\sqrt{\frac{\bar{s}}{\bar{k}}}}} \\ m_4 = e^{\bar{h}_{IL}\sqrt{p\sqrt{\frac{\bar{k}}{\bar{s}}}}} - e^{-\bar{h}_{IL}\sqrt{p\sqrt{\frac{\bar{k}}{\bar{s}}}}} \end{cases} \tag{10}$$

and functions $F$ and $G$ are given by

$$\begin{cases} F = e^{(\bar{x}+\bar{h}_{OL})\sqrt{p\sqrt{\frac{\bar{s}}{\bar{k}}}}} - e^{-(\bar{x}+\bar{h}_{OL})\sqrt{p\sqrt{\frac{\bar{s}}{\bar{k}}}}} \\ G = e^{(\bar{x}-\bar{h}_{IL})\sqrt{p\sqrt{\frac{\bar{k}}{\bar{s}}}}} + e^{-(\bar{x}-\bar{h}_{IL})\sqrt{p\sqrt{\frac{\bar{k}}{\bar{s}}}}} \end{cases} \tag{11}$$

The temperature on the inner wall ($x = h_{IL}$) of the bilayer, denoted as $\bar{T}_{IW}$, thus is given by

$$\overline{T}_{\mathrm{IW}}(\overline{t}) = \overline{T}(\overline{x} = \overline{h}_{\mathrm{IL}}, \overline{t}) = \mathcal{L}^{-1}\left[\frac{4\sqrt{\overline{k}\overline{s}}\left(1 - e^{-\overline{t}_{\mathrm{IM}}p}\right)}{p\left(m_1 m_2 \sqrt{\overline{k}\overline{s}} + m_3 m_4\right)}\right] \tag{12}$$

Given $\overline{k}$, $\overline{s}$, $\overline{t}_{\mathrm{IM}}$, and $\overline{h}_{\mathrm{OL(IL)}}$, the temperature $\overline{T}_{\mathrm{IW}}$ given by Eq. (12) can be calculated numerically for any time $\overline{t}$ with MATLAB (The MathWorks, Inc.), giving rise to the numerical solution to the temporal evolution of $\overline{T}_{\mathrm{IW}}(\overline{t})$.

**3. Results and discussions**

Taking $\overline{k} = 0.1$, $\overline{s} = 10$, $\overline{h}_{\mathrm{OL}} = 0.5$, Fig. 2b shows the calculated evolution of $\overline{T}_{\mathrm{IW}}$ in response to thermal impulses with different durations of $\overline{t}_{\mathrm{IM}} = 0.4, 4, 40$. One can see that $\overline{T}_{\mathrm{IW}}$ exhibits a similar trend of evolution. At $\overline{t} = 0$ when the external impulse is applied, it starts to grow until a moment shortly after the cease of the thermal impulse at $\overline{t} = \overline{t}_{\mathrm{IM}}$. After that, $\overline{T}_{\mathrm{IW}}$ declines gradually to zero as time goes by. The apex of the $\overline{T}_{\mathrm{IW}}$ during this process is denoted by $\overline{T}_{\mathrm{IW}}^{m}$. Fig. 2b shows that if the impulse duration is short, $\overline{T}_{\mathrm{IW}}^{m}$ could be much lower than the temperature applied on the outer wall, implying the considerable thermal resistance of the bilayer against thermal impulse. Therefore, the magnitude of $\overline{T}_{\mathrm{IW}}^{m}$ can be applied to quantify the thermal resistance of a bilayer to given impulse. The lower the $\overline{T}_{\mathrm{IW}}^{m}$, the better the thermal resistance. Recalling the definition of the normalized time in Eq. (2), one can see that $\overline{t}_{\mathrm{IM}}$ is not only proportional to the impulse duration $(t_{\mathrm{IM}})$, but also proportional to the geometric mean of the thermal diffusivities (*i.e.*, $k/s$) of two layers and inversely proportional to $(h_{\mathrm{OL}} + h_{\mathrm{IL}})^2$. This implies that applying materials with lower thermal diffusivities or increasing the overall thickness of the bilayer will also lead to lower $\overline{t}_{\mathrm{IM}}$ and therefore benefits the thermal resistance. A question remains.

How can we maximize the thermal resistance of a bilayer composed of two layers with given materials? In the following, we will investigate the optimization of thermal resistance along two dimensions: layer sequence and volume fraction.

*3.1 Effect of layer sequence on thermal resistance*

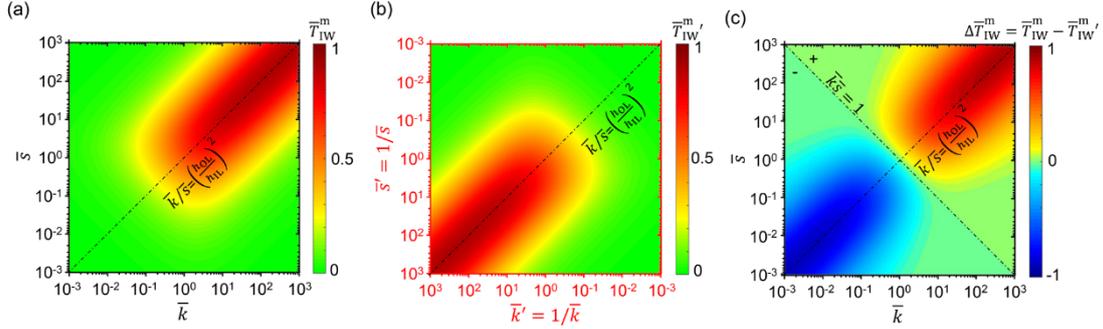

**Fig. 3 (a)** Contour plot of the maximum responsive temperature on the inner wall ($\overline{T}_{IW}^{m}$) with $\overline{k}$ and $\overline{s}$ in the range of $[10^{-3}, 10^{3}]$. **(b)** Contour plot of the maximum responsive temperature on the inner wall of the reciprocal bilayers with a swapped layer sequence ($\overline{T}_{IW}^{m}{}'$). **(c)** Contour plot of the difference between $\overline{T}_{IW}^{m}$ and $\overline{T}_{IW}^{m}{}'$. Here, $\overline{h}_{OL} = 0.5$ and $\overline{t}_{IM} = 0.4$.

The first factor we would like to investigate is the layer sequence in the bilayer. Fig. 3a shows the contour plot of $\overline{T}_{IW}^{m}$ on the $\overline{k}$-$\overline{s}$ plane (logarithmic scale) in the domain of $\overline{k} \in [10^{-3}, 10^{3}]$ and $\overline{s} \in [10^{-3}, 10^{3}]$. For the moment, it is assumed that the OL and IL have the same thickness, namely, $\overline{h}_{OL} = 0.5$. It can be seen that the contour of $\overline{T}_{IW}^{m}$ is symmetric about the line of $\overline{k} = \overline{s}\left(\frac{\overline{h}_{OL}}{\overline{h}_{IL}}\right)^{2}$ (see **Appendix** for elaboration). Elevated $\overline{T}_{IW}^{m}$ occurs as $\overline{k}$ and $\overline{s}$ grow along this line of symmetry. This is the scenario that one should avoid when designing a bilayer for thermal resistance application. The simplest way to reduce $\overline{T}_{IW}^{m}$ is to swap the stacking sequence of the OL and IL. To evaluate the effect of swapping layer sequence on the thermal resistance, the maximum temperature on the inner wall of a bilayer with swapped layer sequence, which is denoted by $\overline{T}_{\text{IW}}^{\text{m}}{'}$, is plotted in Fig. 3b. The difference between $\overline{T}_{\text{IW}}^{\text{m}}$ and $\overline{T}_{\text{IW}}^{\text{m}}{'}$, which is denoted as $\Delta\overline{T}_{\text{IW}}^{\text{m}}$, then is plotted in Fig. 3c, showing the effect of swapping layer sequence on the thermal resistance. It can be seen that $\Delta\overline{T}_{\text{IW}}^{\text{m}}$ is negative when $\overline{k}\overline{s} < 1$ and positive when $\overline{k}\overline{s} > 1$. This implies that for higher thermal resistance two constitutive layers should be placed in such a sequence that the product of $\overline{k}$ and $\overline{s}$ is less than 1. Fig. 3c also indicates that for bilayers with $\overline{k}\overline{s} = 1$, $\overline{T}_{\text{IW}}^{\text{m}}$ is insensitive to the layer sequence. This can be attributed to the intrinsic exchangeability of $\overline{h}_{\text{IL}}^{2}\overline{k}$ and $\overline{h}_{\text{OL}}^{2}\overline{s}$ in the expression of $\overline{T}_{\text{IW}}$ (see **Appendix** for elaboration).

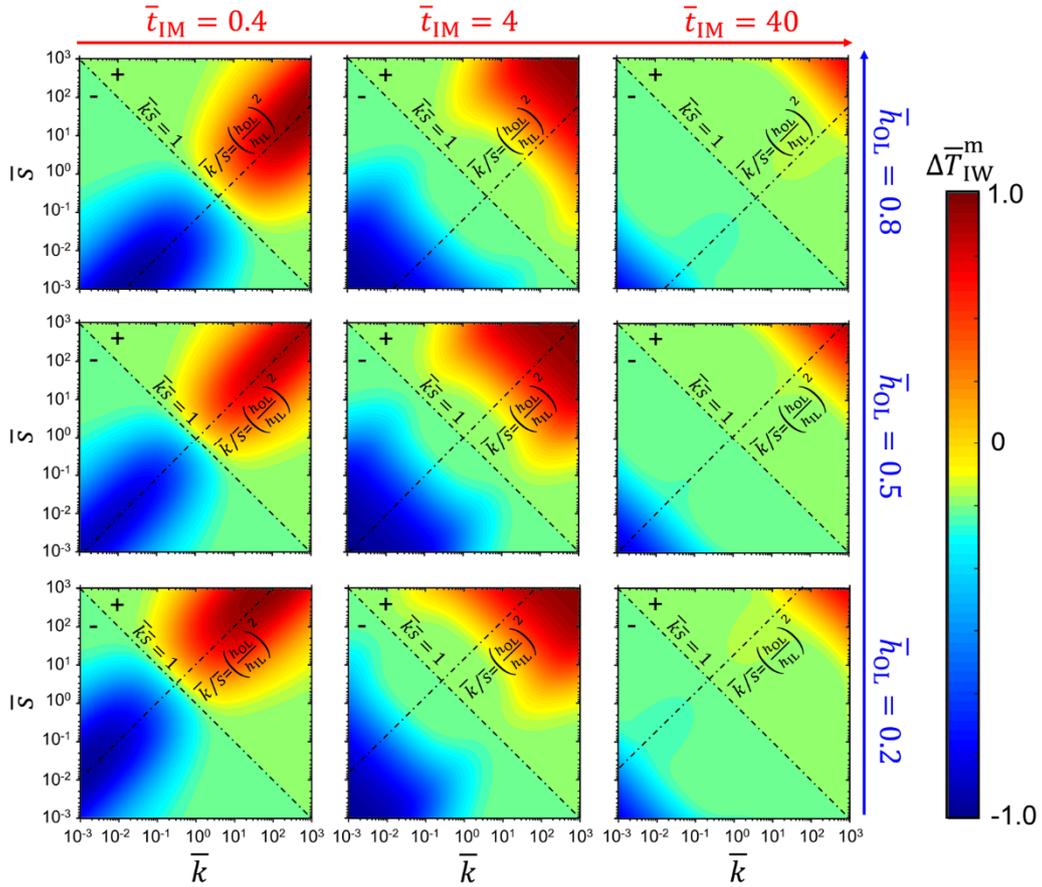

**Fig. 4.** Contour plots of the difference between $\overline{T}_{\text{IW}}^{\text{m}}$ and $\overline{T}_{\text{IW}}^{\text{m}}{'}$ with $\overline{h}_{\text{OL}} = 0.2, 0.5, 0.8$ and $\overline{t}_{\text{IM}} = 0.4, 4, 40$.

The above discussion can be further extended to the bilayers with dissimilar thickness ratio between the OL and IL. Fig. 4 shows the contour plots of $\Delta \overline{T}_{IW}^{m}$ for bilayers with $\overline{h}_{OL} = 0.2, 0.8$. Evidently, $\Delta \overline{T}_{IW}^{m}$ is negative when $\overline{k}\overline{s} < 1$, irrespective of the values of $\overline{h}_{OL}$. Moreover, Fig. 4 indicates that varying of $\overline{h}_{OL}$ causes a translation of the contour of $\Delta \overline{T}_{IW}^{m}$ on the $\overline{k}$-$\overline{s}$ plane. This implies that the thermal resistance of a bilayer can be further optimized by tuning $\overline{h}_{OL}$, as elaborated next.

*3.2 Effect of volume fraction on thermal resistance*

Having determined the optimal layer sequence, the remaining dimension we can manipulate for higher thermal resistance is the volume fractions of two materials, which are equivalent to the thickness fractions $\overline{h}_{OL(IL)}$ in our bilayer model. Consider a bilayer with optimal layer sequence, namely, $\overline{k}\overline{s} < 1$. If $\overline{k} < 1$ and $\overline{s} \geq 1$, the OL has lower thermal conductivity but higher volumetric heat capacity than the IL does. Under this circumstance, having thicker OL and thinner IL benefits the thermal resistance of the bilayer provided that the total thickness is fixed. Therefore, $\overline{T}_{IW}^{m}$ monotonically decreases as $\overline{h}_{OL}$ increases from 0 to 1. In contrast, if $\overline{k} \geq 1$ and $\overline{s} < 1$, the OL has higher thermal conductivity but lower volumetric heat capacity than the IL does. $\overline{T}_{IW}^{m}$ monotonically increases with $\overline{h}_{OL}$. Under this circumstance, the thinner the OL the higher the thermal resistance of the bilayer. To illustrate the dependence of thermal resistance on the thickness fraction in these two cases, we plot the variation of $\overline{T}_{IW}^{m}$ with $\overline{h}_{OL}$ by taking $\overline{k} = 0.1$, $\overline{s} = 1$ and $\overline{k} = 1$, $\overline{s} = 0.1$, as shown in Fig. 5a.

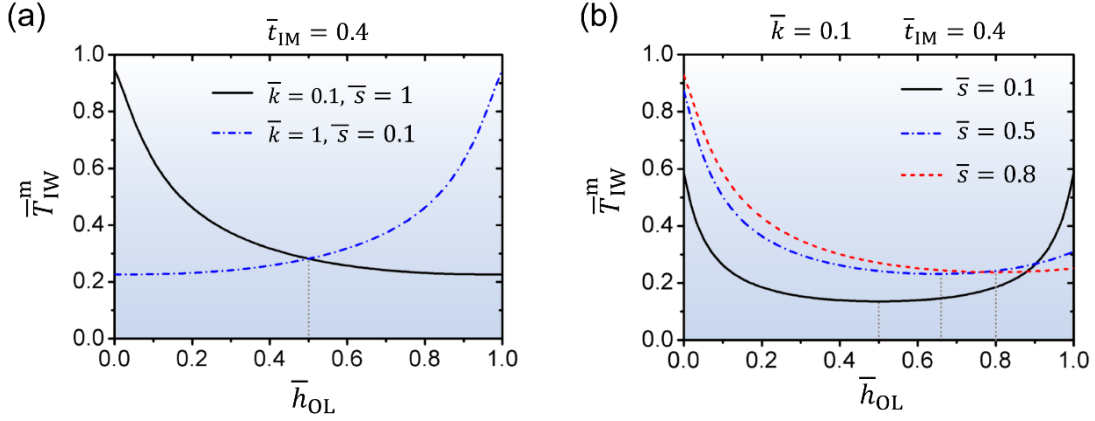

**Fig. 5** The dependence of the maximum temperature experienced by the inner wall ($\overline{T}_{IW}^m$) on the thickness fraction of the OL ($\overline{h}_{OL}$) for **(a)** $\overline{k} = 0.1$, $\overline{s} = 1$ and $\overline{k} = 1$, $\overline{s} = 0.1$, and **(b)** $\overline{k} = 0.1$, $\overline{s} = 0.1, 0.5, 0.8$, respectively.

In addition to the above two scenarios, it might be also possible that the OL in the bilayer is inferior in both lower thermal conductivity and volumetric heat capacity in comparison to the IL, namely $\overline{k} < 1$ and $\overline{s} < 1$. Under this circumstance, $\overline{T}_{IW}^m$ does not exhibit a monotonic dependence on $\overline{h}_{OL}$ due to the competing effects of thermal conductivity and heat capacity on thermal resistance. Instead, there exists an optimal $\overline{h}_{OL}$, at which $\overline{T}_{IW}^m$ is minimized, as illustrated by Fig. 5b. Therefore, the thermal resistance of the bilayer can be maximized by adopting the optimal $\overline{h}_{OL}$, which is dependent on the values of $\overline{k}$ and $\overline{s}$. Recall the multilayer design of the snail shells from the hydrothermal environment. It is of great interest to verify whether nature has adopted such an optimal design for higher thermal resistance.

*3.3 Snail shells from the hydrothermal environment: Optimal design for higher thermal resistance by nature*

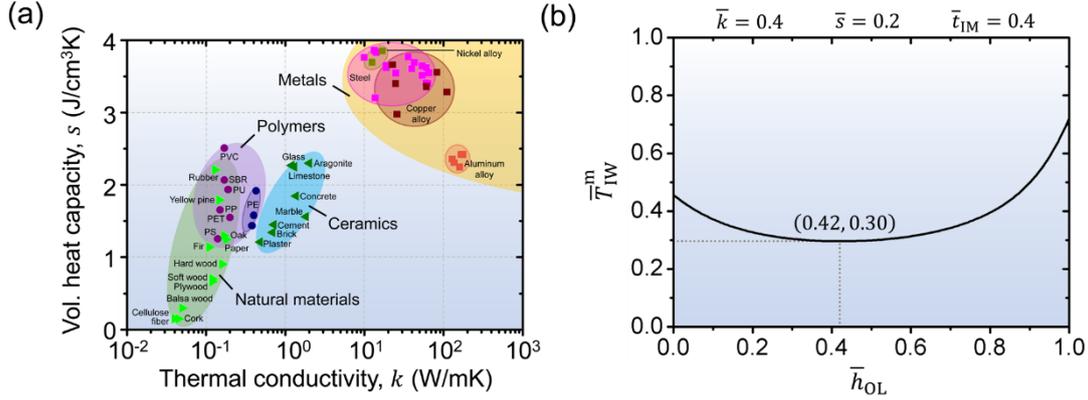

**Fig. 6** (a) Ashby diagram of thermal conductivity versus volumetric heat capacity for typical structural and heat-resistant engineering materials [12-17]. (b) Variation of $\overline{T}_{IW}^{m}$ with $\overline{h}_{OL}$ for a bilayer composed of protein-based material and aragonite.

In our preceding discussion, we have not considered the realistic ranges of thermal conductivity ($k$) and volumetric heat capacity ($s$). Actually, $k$ and $s$ vary in different ranges, as shown by the $k$-$s$ Ashby plot in Fig. 6a. It can be seen that $k$ spans three orders of magnitude from $10^{-1}$ to $10^2$ W m$^{-1}$K$^{-1}$, while $s$ ranges only from 0.1 to 4 J cm$^{-3}$K$^{-1}$. For animals in nature, the materials available for constructing exoskeletal shields are limited, including calcified ceramics and organic materials. For example, most seashells are composed of calcium carbonate, typically aragonite, and protein-based materials. Given these two kinds of materials, how can we design a bilayer with higher thermal resistance?

For aragonite, the typical value of thermal conductivity and volumetric heat capacity are around 2.0 W m$^{-1}$K$^{-1}$[16] and 2.3 J cm$^{-3}$K$^{-1}$[17], respectively. On the other hand, for the organic phase like protein, chitin and other biomacromolecule matters, the typical values of $k$ and $s$ are around 0.8 W m$^{-1}$K$^{-1}$[18] and 0.46 J cm$^{-3}$K$^{-1}$ [19, 20], respectively, which approximately lie in the domain of natural materials in the Ashby plot (Fig. 6a). Apparently, the organic phase is inferior in both thermal conductivity and heat capacity in comparison to the calcified ceramic phase. According to our result in Section 3.1, the organic phase should be placed outside of the ceramic phase in order to achieve a higher thermal resistance, resulting in a bilayer with $\bar{k} = 0.4$ and $\bar{s} = 0.2$. For this bilayer, the thermal resistance can be further optimized by tuning the thickness fraction $\bar{h}_{OL}$ since $\bar{k}\bar{s} < 1$ as indicated by Section 3.2. The calculated variation of $\bar{T}_{IW}^{m}$ with $\bar{h}_{OL}$ for $\bar{k} = 0.4$ and $\bar{s} = 0.2$ is shown in Fig. 6b. The optimal volume thickness fraction is estimated to be $\bar{h}_{OL} \approx 0.42$. This estimation is consistent with the thickness fraction of the organic layer in the shell of *Crysomallon squamiferum*, a snail from deep-sea hydrothermal vent [6]. It is believed that the unique shell structure evolved in these snails is most probably a consequence of adaptation to the extreme temperature variation near the deep-sea hydrothermal vents for higher survivability of the species.

## 4. Conclusions

In this paper, we theoretically studied the effect of structural determining factors, including layer sequence and volume fraction, on the thermal resistance of a bilayer to the external thermal impulse. Based on our results, three practical guidelines for designing bilayers with higher thermal resistance are proposed as follows:

1. To design a bilayer with higher thermal resistance, one should select materials with thermal diffusivity as low as possible. That is, materials with lower thermal conductivity ($k$) and higher volumetric heat capacity ($s$) are preferred.

2. For two layers with distinct thermal properties, their stacking sequence plays an important role in determining the overall thermal resistance of the bilayer. For higher thermal resistance of the bilayer, one should place the two layers in such a sequence that the product of the conductivity ratio and volumetric capacity ratio between the OL and IL is less than 1, namely, $\frac{k_{OL}}{k_{IL}}\frac{s_{OL}}{s_{IL}} < 1$.

3. For a bilayer with an optimized layer sequence, the thermal resistance can be further optimized by tuning the thickness fraction of the layers. If $\frac{k_{OL}}{k_{IL}} < 1$ and $\frac{s_{OL}}{s_{IL}} \geq 1$ (or, alternatively $\frac{k_{OL}}{k_{IL}} \geq 1$ and $\frac{s_{OL}}{s_{IL}} < 1$), thicker OL (or IL) leads to the higher thermal resistance of the bilayer. If $\frac{k_{OL}}{k_{IL}} < 1$ and $\frac{s_{OL}}{s_{IL}} < 1$, there exist optimal thickness fractions, at which the thermal resistance of the bilayer is maximized.

Our findings not only explain the success of the deep-sea snails in surviving the thermal impulses from the hydrothermal vents but also provide a theoretical basis for the design and optimization of thermal barriers in engineering.

**Appendix**

Eq. (12) can be rewritten as

$$\bar{T}_{IW} = \mathcal{L}^{-1}\left[\frac{4\left(1-e^{-\bar{t}_{IM}p}\right)\sqrt{\left(\bar{h}_{IL}^2\bar{k}\right)\cdot\left(\bar{h}_{OL}^2\bar{s}\right)}}{p\left(m_1 m_2\sqrt{\left(\bar{h}_{IL}^2\bar{k}\right)\cdot\left(\bar{h}_{OL}^2\bar{s}\right)}+m_3 m_4 \bar{h}_{OL}\bar{h}_{IL}\right)}\right] \quad (A1)$$

where $m_1$, $m_2$, $m_3$ and $m_4$ are given by

$$\begin{cases} m_1 = e^{-\sqrt{\bar{h}_{OL}\bar{h}_{IL}p}\sqrt{\frac{\bar{h}_{OL}^2\bar{s}}{\bar{h}_{IL}^2\bar{k}}}} + e^{\sqrt{\bar{h}_{OL}\bar{h}_{IL}p}\sqrt{\frac{\bar{h}_{OL}^2\bar{s}}{\bar{h}_{IL}^2\bar{k}}}} \\ m_2 = e^{-\sqrt{\bar{h}_{OL}\bar{h}_{IL}p}\sqrt{\frac{\bar{h}_{IL}^2\bar{k}}{\bar{h}_{OL}^2\bar{s}}}} + e^{\sqrt{\bar{h}_{OL}\bar{h}_{IL}p}\sqrt{\frac{\bar{h}_{IL}^2\bar{k}}{\bar{h}_{OL}^2\bar{s}}}} \\ m_3 = e^{\sqrt{\bar{h}_{OL}\bar{h}_{IL}p}\sqrt{\frac{\bar{h}_{OL}^2\bar{s}}{\bar{h}_{IL}^2\bar{k}}}} - e^{-\sqrt{\bar{h}_{OL}\bar{h}_{IL}p}\sqrt{\frac{\bar{h}_{OL}^2\bar{s}}{\bar{h}_{IL}^2\bar{k}}}} \\ m_4 = e^{\sqrt{\bar{h}_{OL}\bar{h}_{IL}p}\sqrt{\frac{\bar{h}_{IL}^2\bar{k}}{\bar{h}_{OL}^2\bar{s}}}} - e^{-\sqrt{\bar{h}_{OL}\bar{h}_{IL}p}\sqrt{\frac{\bar{h}_{IL}^2\bar{k}}{\bar{h}_{OL}^2\bar{s}}}} \end{cases} \quad (A2)$$

From Eq. (A1) and (A2), it can be easily proved that terms of $\bar{h}_{IL}^2\bar{k}$ and $\bar{h}_{OL}^2\bar{s}$ in the expression of $\bar{T}_{IW}$ are mathematically exchangeable. It is this exchangeability that accounts for the symmetry of the contour of $\bar{T}_{IW}^m$ about the line of $\bar{h}_{IL}^2\bar{k} = \bar{h}_{OL}^2\bar{s}$ or $\bar{k} = \frac{\bar{h}_{OL}^2}{\bar{h}_{IL}^2}\bar{s}$ on the $\bar{k}$-$\bar{s}$ plane. As for the bilayers with $\bar{k}\bar{s} = 1$, the corresponding reciprocal bilayers with swapped layer sequence still satisfy $\bar{k}\bar{s} = 1$. On the logarithmic scale, the line $\bar{k}\bar{s} = 1$ is perpendicular to the symmetry axis $\bar{k} = \frac{\bar{h}_{OL}^2}{\bar{h}_{IL}^2}\bar{s}$. Thus, swapping layer sequence does not affect the values of $\bar{T}_{IW}^m$ for the bilayers with $\bar{k}\bar{s} = 1$.

**Declaration of Competing Interest**

None

**Acknowledgments**

This work was supported by the Research Grant Council of Hong Kong (Grant No.: PolyU152064/15E) and the General Research Fund of Hong Kong Polytechnic University (G-UAHN).